# Group classification of the Sachs equations for a radiating axisymmetric, non-rotating, vacuum space-time.

## Nail H. Ibragimov[1], Ewald J. H. Wessels[2] and George F. R. Ellis[2]

[1] Research Centre ALGA: Advances in Lie Group Analysis
Blekinge Institute of Technology
SE-371 79 Karlskrona, Sweden
[2] Department of Applied Mathematics
University of Cape Town
Cape Town, South Africa

**Abstract.** We carry out a Lie group analysis of the Sachs equations for a time-dependent axisymmetric non-rotating space-time in which the Ricci tensor vanishes. These equations, which are the first two members of the set of Newman-Penrose equations, define the characteristic initial-value problem for the space-time. We find a particular form for the initial data such that these equations admit a Lie symmetry, and so defines a geometrically special class of such spacetimes. These should additionally be of particular physical interest because of this special geometric feature.

PACS numbers: 04.20.Ex, 04.20.-q, 04.30.-w, 02.20.Qs

**Key words:** Initial value problem, Sachs equations, symmetries, equivalence transformations, group classification.

## 1. Introduction

The advent of gravitational wave detectors has increased the need for realistic mathematical models of radiating space-times. The first step in such modeling is the initial value problem for Einstein's field equations. Christodoulou [1] has called this mathematical problem the problem "par excellence" in general relativity.

The usual approach is to choose initial data on a space-like Cauchy slice. The initial data must then satisfy Einstein's equations on that slice (see e.g. [2] for a review).

An alternative is the "characteristic initial value" approach, which uses null geodesics, or "characteristics" as coordinate curves (see e.g. [3] for a review). In this case the initial value function is defined on a null hypersurface. One of the many advantages is the fact that analytic initial data, in the form of the first element $\Psi_0$ of the Weyl spinor, are then free of any constraints, other than conditions that may be imposed on the geometry of the space-time, such as the conditions that are imposed in this paper. However, in the words of Winicour, "This flexibility and control in prescribing initial data has the trade-off of limited experience with prescribing physically realistic characteristic initial data." This comment reflects the fact that initial value functions are usually constructed "by hand" to give suitable limiting values.

In the case of axisymmetric, non-rotating vacuum space-times it has been shown [4] that the initial value problem is defined by the Sachs equations when these are written out in appropriately chosen null coordinates. This pair of equations, originally discovered by Sachs [5], is the first pair of the set of Newman-Penrose equations [6] for the space-time. With a coordinate system based on a set of null geodesics emanating from a point on the axis of symmetry, they are defined on an outgoing null geodesic and relate two of the metric functions, which are labeled Q(u,r,θ) and q(u,r,θ) to the initial-value function ψ₀(u,r,θ). Here u is a time-like coordinate that labels successive null hypersurfaces, r is a space-like affine parameter that measures distance along the null geodesics, and θ is an angular elevation coordinate. The fourth coordinate, φ, is an azimuth and does not appear in these functions because of the rotational symmetry that is assumed to exist about the axis. Along a single null geodesic u, θ and φ are all constant and only r varies. With the symmetries under discussion these equations take the following form when the Ricci tensor vanishes:

$$\frac{\partial^2 q}{\partial r^2} + \Psi_0 \, q = 0 \quad \text{and} \quad \frac{\partial^2 Q}{\partial r^2} - \Psi_0 \, Q = 0 \, . \qquad (1.1)$$

The unknowns q and Q are the $d\theta^2$ metric function and the $d\phi^2$ metric function respectively (see the appendix). The function $\Psi_0$ is the first element of the Weyl spinor.

Penrose [7] has shown that $\Psi_0$ is not determined by the field equations but can be specified independently on an initial null hypersurface without any constraints (other than constraints that may derive from symmetry conditions imposed on the geometry.) The equations can therefore be treated as a closed set of ordinary differential equations that determine the r dependence of q and Q once the character of the space-time has been prescribed by the choice of $\Psi_0$ on the initial null hypersurface.

The Sachs equations (1.1) are underdetermined with respect to a solution, since the two equations contain three unknown variables. However, it has been shown [4] that group theory can provide a third constraint on the problem, in the form of a particular non-trivial Lie symmetry that can be imposed on the equations. If this symmetry (which is the only non-trivial Lie symmetry that the equations can admit for any possible form of the function $\Psi_0$) is imposed they have the following unique general solution:

$$\Psi_0\big|_{u=0} = \frac{a_0}{\left(b_2 r^2 + b_1 r + b_0\right)^2} \qquad (1.2)$$

where $a_0$ and the $b_i$ are unknown functions of θ.

The four undetermined functions of integration appear to allow sufficient latitude to cover the variety of matter distributions that belong to the class of non-rotating axisymmetric space-times (note, as remarked before, that the equations under consideration cover only the vacuum portion of such space-times where the Ricci tensor vanishes). In particular, they cover the spherically symmetric (Schwarzschild) case, in which case $a_0$, and therefore $\Psi_0$, vanish identically.

One particular choice of values for the functions of integration (when $b_1 = 0$) appears to describe the vacuum portion of a space-time corresponding to a pair of identical non-rotating massive bodies moving along an axis of symmetry towards a head-on collision with the origin at the centre of mass The full set of Newman-Penrose equations can be solved analytically for the elements of the metric tensor, in this case, in the limit when $\theta \to 0$, i.e. on the axis of symmetry [4]. The solution is regular on the axis unlike previous axisymmetric solutions, such as the Weyl solution, which contain conical singularities on the axis. It passes other reasonable tests, which suggest that the Lie symmetry solution approximates to a single Schwarzschild solution and a pair of Schwarzschild solutions respectively when a parameter, which can be identified approximately with the separation of the two massive bodies, is small or large.

Also, the "tidal force" term of the Weyl curvature on the axis, in the weak field limit, approximates to the curvature that can be calculated from Newton's laws for such a case.

These features strongly suggest that the mathematical property of the equations that produces this result is a reflection of an underlying interesting physical property of non-rotating axisymmetric relativistic space-time manifolds and that it is not just a mathematical curiosity [4].

The details of the Lie group analysis that produces the result (1.2) have not previously been published. It is the purpose of this paper to give these details and also to extend the analysis by finding the group of equivalence transformations of the Sachs equations (1.1).

## 2. Symmetries and the Classifying relation

Consider the system (1.1) of the Sachs equations, which can be treated as a pair of ordinary differential equations. We denote differentiation with respect to r by primes, drop the subscript from $\Psi_0$, and write the equations in the following form:

$$q'' + \Psi(r)q = 0, \qquad Q'' - \Psi(r)Q = 0 \qquad (2.1)$$

Note that each of the two equations, considered separately, admits an 8-dimensional Lie algebra. It would be erroneous, however, to conclude that the simultaneous system of the two equations should admit a 16-dimensional Lie Algebra (cf. [8], Section 9.3.3).

The Sachs equations (2.1) are linear and homogeneous. Therefore they admit the six-parameter group composed of the dilations in q and Q and the usual superposition principle consisting of addition to q and Q of the arbitrary solutions k(r) and l(r) of the first and second equation (2.1) respectively. In other words, the Sachs equations (2.1) with an arbitrary potential $\Psi(r)$ admit:

$$X_1 = q\frac{\partial}{\partial q}, \quad X_2 = Q\frac{\partial}{\partial Q}, \quad X_3 = k(r)\frac{\partial}{\partial q}, \quad X_4 = l(r)\frac{\partial}{\partial Q} \qquad (2.2)$$

where k(r) and l(r) solve the Sachs equations

$$k''(r) + \Psi(r)k(r) = 0 \qquad l''(r) - \Psi(r)l(r) = 0$$

Since the general solution of the latter equations depends on 4 arbitrary constants the operators (2.2) span a six-dimensional Lie algebra.

We look for the general admissible infinitesimal generators

$$X = \tau(r,q,Q)\frac{\partial}{\partial r} + \xi(r,q,Q)\frac{\partial}{\partial q} + \eta(r,q,Q)\frac{\partial}{\partial Q} \qquad (2.3)$$

The determining equations are written

$$X[q'' + \Psi(r)q]_{(2.1)} \equiv \zeta_2^1\big|_{(2.1)} + \xi\Psi(r) + q\tau\Psi'(r) = 0 \qquad (2.4)$$

$$X[Q'' - \Psi(r)Q]_{(2.1)} \equiv \varsigma_2^2\big|_{(2.1)} - \eta\Psi(r) - Q\tau\Psi'(r) = 0 \qquad (2.5)$$

where $\big|_{(2.1)}$ means that $q''$ and $Q''$ should be replaced by $-\Psi(r)q$ and $\Psi(r)Q$ respectively according to the system of equations (2.1).

The quantities $\varsigma_2^1$ and $\varsigma_2^2$ are obtained by the usual prolongation procedure (see any book on Lie group analysis, e.g. [9], [10], [8], [11]), namely by the first prolongation

$$\begin{aligned}\varsigma_1^1 &= D_r(\xi) - q'D_r(\tau) = \xi_r + q'\xi_q + Q'\xi_Q - q'(\tau_r + q'\tau_q + Q'\tau_Q) \\ \varsigma_1^2 &= D_r(\eta) - Q'D_r(\tau) = \eta_r + q'\eta_q + Q'\eta_Q - Q'(\tau_r + q'\tau_q + Q'\tau_Q)\end{aligned} \qquad (2.6)$$

and the second prolongation

$$\varsigma_2^1 = D_r(\varsigma_1^1) - q''D_r(\tau), \qquad \varsigma_2^2 = D_r(\varsigma_1^2) - Q''D_r(\tau) \qquad (2.7)$$

respectively.

Upon substituting the expression (2.7) for $\varsigma_2^1$, (2.4) takes the following form:

$$\begin{aligned}&\xi_{rr} + (2\xi_{rq} - \tau_{rr})q' + 2\xi_{rQ}Q' + (\xi_{qq} - 2\tau_{rq})q'^2 + 2(\xi_{qQ} - \tau_{rQ})q'Q' + \xi_{QQ}Q'^2 - \tau_{qq}q'^3 \\ &- 2\tau_{qQ}q'^2 Q' - \tau_{QQ}q'Q'^2 + Q(\xi_Q - \tau_Q q')\Psi(r) - q(\xi_q - 2\tau_r - 3\tau_q q' - 2\tau_Q Q')\Psi(r) \\ &+ \xi\Psi(r) + q\tau\Psi'(r) = 0\end{aligned} \qquad (2.8)$$

We collect here the like terms and annul the coefficients of different powers of $q'$ and $Q'$. The coefficients for the cubic terms $q'^3, q'^2 Q'$ and $q'Q'^2$ yield:

$$\tau_{qq} = 0, \quad \tau_{qQ} = 0, \quad \tau_{QQ} = 0,$$

whence

$$\tau = a(r)q + b(r)Q + c(r) \qquad (2.9)$$

Furthermore, the coefficients for the quadratic terms $Q'^2, q'Q'$, and $q'^2$ yield:

$$\xi_{QQ} = 0, \quad \xi_{qQ} - \tau_{rQ} = 0, \quad \xi_{qq} - 2\tau_{rq} = 0$$

whence, invoking the expression (2.9) for $\tau$ and using (2.2) one obtains:

$$\xi = a'(r)q^2 + b'(r)qQ + A(r)q + m(r)Q + k(r)$$

Annulling the coefficients for $Q'$ in (2.8) one obtains:

$$\xi_{rQ} + q\Psi(r)\tau_Q \equiv [b''(r) + \Psi(r)b(r)]q + m'(r) = 0$$

whence $m'(r) = 0$ and

$$b''(r) + \Psi(r) b(r) = 0 \tag{2.10}$$

Thus $\xi$ has the form

$$\xi = a'(r) q^2 + b'(r) q Q + A(r) q + m Q + k(r), \quad m = \text{const.} \tag{2.11}$$

Likewise, inspecting (2.5), by using the quadratic terms in $q', Q'$ and the term in $q'$ we arrive at the following equations:

$$a''(r) - \Psi(r) a(r) = 0 \tag{2.12}$$

and

$$\eta = a'(r) q Q + b'(r) Q^2 + B(r) Q + s q + l(r), \quad s = \text{const.} \tag{2.13}$$

We return now to (2.8) and annul there the coefficient for $q'$ and the term free of the derivatives $q'$ and $Q'$ to obtain:

$$2\xi_{rq} - \tau_{rr} + [3q\tau_q - Q\tau_Q]\Psi(r) = 0 \tag{2.14}$$

and

$$\xi_{rr} + [\xi + Q\xi_Q - q\xi_q + 2q\tau_r]\Psi(r) + q\tau\Psi'(r) = 0 \tag{2.15}$$

respectively. Upon substituting the expression (2.9) for $\tau$ and the expression (2.11) for $\xi$, (2.14) becomes:

$$3q[a''(r) + a(r)\Psi(r)] + Q[b''(r) - b(r)\Psi(r)] + 2A'(r) - c''(r) = 0$$

whence, taking into account (2.10) and (2.12) and assuming $\Psi(r) \neq 0$, we have $a(r) = b(r) = 0$ and

$$2A'(r) = c''(r) \tag{2.16}$$

Equation (2.15) now takes the form

$$[(A''(r) + 2c'(r)\Psi(r) + c(r)\Psi'(r)]q + 2mQ\Psi(r) + k''(r) + k(r)\Psi(r) = 0$$

If $\Psi(r) \neq 0$, it follows from this equation that $m = 0$,

$$A''(r) + 2c'(r)\Psi(r) + c(r)\Psi'(r) = 0 \tag{2.17}$$

and

$$k''(r) + k(r)\Psi(r) = 0 \tag{2.18}$$

Similar calculations with (2.5) provide the equation $s = 0$ together with the following counterparts of (2.16), (2.17), and (2.18):

$$2B'(r) = c''(r), \quad B''(r) - 2c'(r)\Psi(r) - c(r)\Psi'(r) = 0, \quad l''(r) - l(r)\Psi(r) = 0 \tag{2.19}$$

In consequence, expressions (2.9), (2.11) and (2.13) reduce to the following:

$$\tau = c(r), \quad \xi = A(r)q + k(r), \quad \eta = B(r)Q + l(r) \tag{2.20}$$

Equations (2.16) and (2.17), together with the first and second equations (2.19), yield the following two simultaneous equations:

$$\tfrac{1}{2}c'''(r) + 2c'(r)\Psi(r) + c(r)\Psi'(r) = 0, \quad \tfrac{1}{2}c'''(r) - 2c'(r)\Psi(r) - c(r)\Psi'(r) = 0$$

whence $c'''(r) = 0$, and

$$2c'(r)\Psi(r) + c(r)\Psi'(r) = 0 \tag{2.21}$$

Equation (2.21) involves the potential $\Psi(r)$ and is called a *classifying relation*. Thus, the function $c(r)$ is at most quadratic in r,

$$c(r) = C_1 r^2 + C_2 r + C_3, \quad C_i = \text{const.} \tag{2.22}$$

Furthermore, the classifying relation (2.21) written in the form

$$\frac{\Psi'(r)}{\Psi(r)} = -2\frac{c'(r)}{c(r)}$$

yields $\Psi(r) = C_4/c^2(r)$, provided that $c(r) \neq 0$. On the other hand, if $c(r) = 0$, it follows from (2.16) and the first equation (2.19) that A = const., B = const., and hence the Sachs equations (2.1) have only the self-evident infinitesimal symmetries (2.2). Thus, invoking (2.16), (2.22) and the first equation (2.19) we arrive at the following statement.

**Theorem 1:** The Sachs equations (2.1) admit, along with the trivial infinitesimal symmetries (2.2), a nontrivial Lie algebra if, and only if, the potential $\Psi(r)$ has the following form:

$$\Psi(r) = \frac{\delta}{\left(\alpha r^2 + \beta r + \gamma\right)^2}, \tag{2.23}$$

where $\alpha,\beta,\gamma$ and $\delta$ are arbitrary constants, $\delta \neq 0$.

This is the result first reported by Wessels [4] in the form (1.2) where, in the context of the Newman-Penrose equations, the "constants" are functions of $\theta$ defined on the initial null hypersurface.

The infinitesimal symmetries (2.3) for the equations (2.1) with the potential (2.23) have the coordinates

$$\begin{aligned}
\tau &= C_1 r^2 + C_2 r + C_3 \\
\xi &= (C_1 r + C_4)q + k(r), \quad k'' + k\Psi = 0, \\
\eta &= (C_1 r + C_5)Q + l(r), \quad l'' - l\Psi = 0,
\end{aligned} \tag{2.24}$$

Where $C_i$ = const. The constants $C_1$, $C_2$, and $C_3$ are connected with the potential (2.23) by (2.21):

$$2(2C_1r + C_2)\Psi + (C_1r^2 + C_2r + C_3)\Psi' = 0 \qquad (2.25)$$

If $\alpha \neq 0$, (2.25) yields the result $C_1 \neq 0$, otherwise $C_2 = C_3 = 0$ and hence one has only the trivial symmetries (2.2). Therefore, we let $C_1 = \alpha$ and obtain from (2.25) the result that $C_2 = \beta$ and $C_3 = \gamma$. Thus equations (2.1) with the potential (2.23), where $\alpha \neq 0$, admit the following non-trivial symmetry [4]:

$$X = (\alpha r^2 + \beta r + \gamma)\frac{\partial}{\partial r} + \alpha r q \frac{\partial}{\partial q} + \alpha r Q \frac{\partial}{\partial Q}. \qquad (2.26)$$

In order to complete the group classification, we need to single out all equivalent potentials (2.23). First, we have to find the equivalence transformations of the system of Sachs equations (2.1).

## 3. Equivalence transformations

An equivalence transformation is a smooth, invertible transformation of the dependent and independent variables that leaves the form of the equations unchanged, except possibly for the form of an unspecified function appearing in the equations, which does not have to be the same before and after the transformation. Thus an equivalence transformation of the system (1.1) is an invertible transformation of the variables r, q, Q:

$$\bar{r} = f(r, q, Q), \quad \bar{q} = g(r, q, Q) \quad \overline{Q} = h(r, q, Q)$$

mapping the system (2.1) into a system of the same form,

$$\frac{d^2\bar{q}}{d\bar{r}^2} + \overline{\Psi}(\bar{r})\bar{q} = 0, \qquad \frac{d^2\overline{Q}}{d\bar{r}^2} - \overline{\Psi}(\bar{r})\overline{Q} = 0$$

where the form of the transformed function $\overline{\Psi}(\bar{r})$ can, in general, be different from the form of the original function $\Psi(r)$. An equivalence transformation can also be regarded as a mapping of the variables (r, q, Q, $\Psi$) into variables $(\bar{r}, \bar{q}, \overline{Q}, \overline{\Psi})$ leaving invariant the system (2.1) written in the "extended form":

$$q'' + q\Psi = 0, \quad Q'' - Q\Psi = 0, \quad \Psi_q = 0, \quad \Psi_Q = 0 \qquad (3.1)$$

Using the latter definition and assuming that the transformed variables $\bar{r}, \bar{q},$ and $\overline{Q}$ involve only r, q, and Q, but not $\Psi$, one can calculate in the usual way the generators of the continuous group of equivalence transformations in the form

$$Y = \tau(r,q,Q)\frac{\partial}{\partial r} + \xi(r,q,Q)\frac{\partial}{\partial q} + \eta(r,q,Q)\frac{\partial}{\partial Q} + \mu(r,q,Q,\Psi)\frac{\partial}{\partial \Psi} \qquad (3.2)$$

Then one can apply Lie's infinitesimal technique by using the prolongation of the operator Y to the derivatives involved in the extended system (3.1) as follows (see [9] and e.g. [12]):

$$\overline{Y} = \tau\frac{\partial}{\partial r} + \xi\frac{\partial}{\partial q} + \eta\frac{\partial}{\partial Q} + \mu\frac{\partial}{\partial \Psi} + \varsigma_2^1\frac{\partial}{\partial q''} + \varsigma_2^2\frac{\partial}{\partial Q''} + \omega_1\frac{\partial}{\partial \Psi_q} + \omega_2\frac{\partial}{\partial \Psi_Q} \tag{3.3}$$

The invariance test for equations (3.1) requires that the following system of *determining equations* holds:

$$\varsigma_2^1\big|_{(3.1)} + \xi\Psi + q\mu = 0 \qquad \varsigma_2^2\big|_{(3.1)} - \eta\Psi - Q\mu = 0 \tag{3.4}$$

$$\omega_1\big|_{(3.1)} = 0 \qquad \omega_2\big|_{(3.1)} = 0 \tag{3.5}$$

Here $\varsigma_2^1$ and $\varsigma_2^2$ are given by the previous prolongation formulae (2.6) and (2.7), whereas $\omega_1$ and $\omega_2$ are determined by

$$\begin{aligned}\omega_1 &= \tilde{D}_q(\mu) - \Psi_q\tilde{D}_q(\xi) - \Psi_Q\tilde{D}_q(\eta) - \Psi_r\tilde{D}_q(\tau),\\ \omega_2 &= \tilde{D}_Q(\mu) - \Psi_q\tilde{D}_Q(\xi) - \Psi_Q\tilde{D}_Q(\eta) - \Psi_r\tilde{D}_Q(\tau)\end{aligned} \tag{3.6}$$

where

$$\tilde{D}_q = \frac{\partial}{\partial q} + \Psi_q\frac{\partial}{\partial \Psi}, \qquad \tilde{D}_Q = \frac{\partial}{\partial Q} + \Psi_Q\frac{\partial}{\partial \Psi} \tag{3.7}$$

are the "new" total differentiations for the extended system (3.1).

The restriction to the equations (3.1) means, in particular, that we set $\Psi_q = \Psi_Q = 0$. The expressions (3.7) and (3.6) take the form:

$$\tilde{D}_q = \frac{\partial}{\partial q}, \qquad \omega_1 = \mu_q - \Psi_r\tau_q$$

$$\tilde{D}_Q = \frac{\partial}{\partial Q}, \qquad \omega_2 = \mu_Q - \Psi_r\tau_Q$$

Let us begin with the equations (3.5):

$$\omega_1 \equiv \mu_q - \Psi_r\tau_q = 0, \qquad \omega_2 \equiv \mu_Q - \psi_r\tau_Q = 0$$

Since $\Psi$ and hence $\Psi_r$ are arbitrary functions, the above equations yield:

$$\tau_q = 0, \quad \mu_q = 0, \quad \tau_Q = 0, \quad \mu_Q = 0.$$

Thus the operator (3.2) reduces to the form:

$$Y = \tau(r)\frac{\partial}{\partial r} + \xi(r,q,Q)\frac{\partial}{\partial q} + \eta(r,q,Q)\frac{\partial}{\partial Q} + \mu(r,\Psi)\frac{\partial}{\partial \Psi} \tag{3.8}$$

Comparing equations (3.4) with the determining equations for the symmetries we notice that equations (3.4) are obtained from (2.4) and (2.5) merely by replacing $\tau \Psi'(r)$ by $\mu$. Hence we can apply the calculations of the previous section and arrive at the equations (2.24). We note that since $\Psi$ is regarded now as an arbitrary variable, the equations $k''(r) + k(r)\Psi(r) = 0$ and $l''(r) - l(r)\Psi(r) = 0$ yield $k(r) = 0$ and $l(r) = 0$, respectively. Furthermore we make the above mentioned replacement $\tau \Psi'(r) \mapsto \mu$ in (2.25).

Summing up, we obtain the following coordinates of the equivalence generator (3.8):

$$\begin{aligned}
\tau &= C_1 r^2 + C_2 r + C_3, \\
\xi &= (C_1 r + C_4) q \\
\eta &= (C_1 r + C_5) Q \\
\mu &= -(4 C_1 r + 2 C_2) \Psi
\end{aligned} \qquad (3.9)$$

Thus, the system (2.1) has the five-dimensional equivalence Lie algebra spanned by the following equivalence generators:

$$Y_1 = q \frac{\partial}{\partial q}, \quad Y_2 = Q \frac{\partial}{\partial Q}, \quad Y_3 = \frac{\partial}{\partial r}, \quad Y_4 = r \frac{\partial}{\partial r} - 2\Psi \frac{\partial}{\partial \Psi},$$

$$Y_5 = r^2 \frac{\partial}{\partial r} + rq \frac{\partial}{\partial q} + rQ \frac{\partial}{\partial Q} - 4r\Psi \frac{\partial}{\partial \Psi} \qquad (3.10)$$

The operators $Y_1$, $Y_2$, and $Y_3$ generate the groups of dilations in q and Q and the translations in r respectively. The operator $Y_4$ generates the group of dilations in r together with a coupled dilation of $\Psi$, i.e. $\bar{r} = ar$, $\bar{\Psi} = \frac{\Psi}{a^2}$. The operator $Y_5$ generates the group

$$\bar{r} = \frac{r}{1-ar}, \quad \bar{q} = \frac{q}{1-ar}, \quad \bar{Q} = \frac{Q}{1-ar}, \quad \bar{\Psi} = (1-ar)^4 \Psi \qquad (3.11)$$

## 4. Classification

Let us find all the potentials (2.23) that are equivalent to the constant potential

$$\Psi = C, \qquad C \neq 0 \qquad (4.1)$$

In order to find them, we substitute $\bar{\Psi} = \delta$, where $\delta$ is a constant, in the equivalence transformation (3.11), $(1-ar)^4 \Psi = \delta$, and obtain

$$\Psi(r) = \frac{\delta}{(1-ar)^4} \equiv \frac{\delta}{(a^2 r^2 - 2ar + 1)^2} \qquad (4.2)$$

This is a potential of the form (2.23) with $\alpha = a^2$, $\beta = -2a$, and $\gamma = 1$. For this potential, the expression $\alpha r^2 + \beta r + \gamma$ has the vanishing discriminant

$$\Delta = \beta^2 - 4\alpha\gamma, \tag{4.3}$$

i.e. the equation $\alpha r^2 + \beta r + \gamma = 0$ has two equal roots. One can readily verify that the remaining equivalence transformations, i.e. the dilations and the translation of r leave invariant the equation $\Delta = 0$. Therefore, we can obtain from $\overline{\Psi}(\bar{r}) = C$ all potentials (2.23) with the vanishing discriminant (4.3). Vice versa, all potentials (2.23) with the vanishing discriminant (4.3) are equivalent to the constant potential, since the equivalence transformations are invertible. For the constant potential (4.1), (2.25) yields $C_1 = C_2 = 0$. Hence, the Sachs equations (2.1) with the constant potential (4.1) have, along with the six trivial symmetries (2.2), the additional symmetry

$$X_7 = \frac{\partial}{\partial r} \tag{4.4}$$

Likewise, one readily finds the potentials that are equivalent to

$$\Psi(r) = \frac{C}{r^2}, \qquad C \neq 0 \tag{4.5}$$

Namely, substituting $\overline{\Psi} = \frac{\delta}{\bar{r}^2}$ in the transformation (3.11),

$$(1 - ar)^4 \Psi = \frac{(1-ar)^2}{r^2}\delta$$

one obtains:

$$\Psi(r) = \frac{\delta}{r^2(1-ar)^2} \equiv \frac{\delta}{(ar^2 - r)^2}$$

This is a potential of the form (2.23) with $\alpha = a$, $\beta = -1$, and $\gamma = 0$. It has the positive discriminant $\Delta > 0$. In other words, the equation $\alpha r^2 + \beta r + \gamma = 0$ has two real roots. As above, we conclude that all potentials (2.23) with the positive discriminant (4.3) are equivalent to the potential (4.5), since the equivalence transformations are invertible and do not change the sign of the discriminant (4.3). For the potential (4.5), (2.25) yields $C_1 = C_3 = 0$. Hence, the equations (2.1) with the potential (4.5) have the following additional symmetry (written upon subtracting the operators $X_1$ and $X_2$):

$$X_7 = r\frac{\partial}{\partial r} \tag{4.6}$$

Furthermore, one can readily verify that the potentials (2.23) with the negative discriminant (4.3), $\Delta < 0$, are equivalent to the potential

$$\Psi(r) = \frac{C}{(r^2+1)^2}, \qquad C \neq 0 \tag{4.7}$$

The equations (2.1) with the potential (4.7) have the additional symmetry (2.26) with $\alpha = 1$, $\beta = 0$ and $\gamma = 1$:

$$X_7 = (1+r^2)\frac{\partial}{\partial r} + rq\frac{\partial}{\partial q} + rQ\frac{\partial}{\partial Q} \tag{4.8}$$

Thus we have proved the following theorem.

**Theorem 2.** The Sachs equations (2.1) with the potential (2.23),

$$\Psi(r) = \frac{\delta}{\left(\alpha r^2 + \beta r + \gamma\right)^2},$$

can be reduced by a proper equivalence transformation to the equation with the standard potential (4.1), (4.5) or (4.7). Specifically,, we can set

$$\begin{aligned} \Psi &= C, & \text{if} \quad \Delta &= 0 \\ \Psi &= \frac{C}{r^2}, & \text{if} \quad \Delta &> 0 \\ \Psi &= \frac{C}{(r^2+1)^2} & \text{if} \quad \Delta &< 0 \end{aligned} \tag{4.9}$$

where $\Delta = \beta^2 - 4\alpha\gamma$.

Therefore it suffices to solve the Sachs equations (2.1) with the standard potentials (4.9).

**Example.** Let us obtain the solution of equations (2.1) with the potential (4.2) where we let $C = \omega^2$:

$$q'' + \frac{\omega^2 q}{(a^2 r^2 - 2ar + 1)^2} = 0, \qquad Q'' - \frac{\omega^2 Q}{(a^2 r^2 - 2ar + 1)^2} = 0 \tag{4.10}$$

In the new variables $\bar{q}$, $\overline{Q}$, and $\bar{r}$ obtained by the equivalence transformations (3.11), the system (4.10) takes the form (cf. (4.1))

$$\bar{q}'' + \omega^2 \bar{q} = 0, \qquad \overline{Q}'' - \omega^2 \overline{Q} = 0$$

Whence

$$\bar{q} = A_1 \cos(\omega \bar{r}) + B_1 \sin(\omega \bar{r}), \qquad \overline{Q} = A_2 e^{\omega \bar{r}} + B_2 e^{-\omega \bar{r}}.$$

According to (3.11) we substitute here

$$\bar{r} = \frac{r}{1-ar},$$

insert the expression obtained for $\bar{q}$ and $\bar{Q}$ into

$$q = (1 - ar)\bar{q}, \quad Q = (1 - ar)\bar{Q}$$

and arrive at the following general solution of the system (4.10):

$$q = (1 - ar)\left[A_1 \cos\left(\frac{\omega r}{1 - ar}\right) + B_1 \sin\left(\frac{\omega r}{1 - ar}\right)\right]$$

$$Q = (1 - ar)\left[A_2 \exp\left(\frac{\omega r}{1 - ar}\right) + B_2 \exp\left(-\frac{\omega r}{1 - ar}\right)\right] \quad (4.11)$$

We recall that the above results were obtained under the assumption that $\Psi \neq 0$. Therefore, to complete the group classification of the systems admitting more general groups than those generated by (2.1), it remains to provide the symmetries of equations (2.1) with the vanishing potential $\Psi$:

$$q'' = 0, \quad Q'' = 0. \quad (4.12)$$

This system admits the 15-dimensional Lie algebra spanned by the following generators (cf. [8] Section 9.3.3):

$$\begin{aligned}
&X_1 = \frac{\partial}{\partial r}, \quad X_2 = \frac{\partial}{\partial q}, \quad X_3 = \frac{\partial}{\partial Q}, \quad X_4 = r\frac{\partial}{\partial r}, \quad X_5 = q\frac{\partial}{\partial r}, \\
&X_6 = Q\frac{\partial}{\partial r}, \quad X_7 = r\frac{\partial}{\partial q}, \quad X_8 = q\frac{\partial}{\partial q}, \quad X_9 = Q\frac{\partial}{\partial q}, \quad X_{10} = r\frac{\partial}{\partial Q}, \\
&X_{11} = q\frac{\partial}{\partial Q}, \quad X_{12} = Q\frac{\partial}{\partial Q}, \quad X_{13} = r^2\frac{\partial}{\partial r} + rq\frac{\partial}{\partial q} + rQ\frac{\partial}{\partial Q}, \\
&X_{14} = rq\frac{\partial}{\partial r} + q^2\frac{\partial}{\partial q} + qQ\frac{\partial}{\partial Q}, \quad X_{15} = rQ\frac{\partial}{\partial r} + qQ\frac{\partial}{\partial q} + Q^2\frac{\partial}{\partial Q}
\end{aligned} \quad (4.13)$$

## 5. Discussion and Conclusions

We start with a discussion of the equivalence transformations derived in section 3.

Since the fundamental postulate of the theory of special relativity is that space-time is a differentiable manifold endowed with a Lorentz group structure, and the Newman-Penrose equations assume that this structure is preserved in the fibre bundle of tangent spaces over the points of a curved space-time, it should not be a surprise that the equivalence transformations of equations (1.1) include the elements of the group SL (2R), the special linear group over the real numbers of order 2: This group is just the group of Lorentz transformations corresponding to a boost in the r - θ plane.

In the coordinates employed in the derivation of equations (1.1) the imaginary axis lies in the direction $\frac{\partial}{\partial \phi}$. It follows that, while the general Lorentz group SL (2C) includes the sub-group of boosts across the direction of the axis of symmetry, the group SL (2R) excludes this sub-group. (For a discussion of the relation between the group SL (2C) and the Lorentz transformations see [13] chapter 1 or [14] chapter 13.) Boosts in the direction of the axis of symmetry, which preserve the axisymmetric character of the space-time and of the coordinates, and therefore must preserve the form of the Sachs equations, belong to the group SL (2R).

The translations in the parameter r that displace the origin of the coordinate system along the axis of symmetry must similarly preserve the form of the Sachs equations.

For the remaining equivalence transformations, the group of translations in the parameter r, in the general case when this displaces the origin away from the axis of symmetry, and boosts in the r - θ plane that are not oriented in the direction of the axis of symmetry, the situation is more complicated. Both these groups of transformations destroy the axisymmetric nature of the coordinate system (though not of the manifold, provided the cross-axis component of the boost vanishes when r = 0). In general, therefore, such transformations can be expected to change the form of the field equations.

However, it should be borne in mind that the Sachs equations, considered in isolation, are defined on a single null geodesic {u,θ,φ = const} emanating from a point on the axis of symmetry. Neither a displacement of the origin along such a coordinate curve, nor a boost in the r - θ plane changes the direction $\frac{\partial}{\partial \phi}$ on the curve itself, where this direction remains aligned to the Killing field. The symmetry of the metric tensor under the reflection $d\phi \to -d\phi$ is therefore preserved in the transformed coordinates and the off-diagonal elements of the fourth row and the fourth column of the metric tensor must still vanish in the new coordinates. As a result the metric tensor on the single null geodesic in question retains the general form that leads to the form of the Sachs equations (see the appendix) and these equations must therefore retain their form after the transformation, with some initial value function $\Psi_0$ (not necessarily the same function). This result, of course, does not apply to the remainder of the Newman-Penrose equations.

We have therefore arrived at the main conclusion of this paper: Any initial value function $\Psi_0$ that is an exact representation of an axisymmetric, non-rotating space-time manifold in which the Newman-Penrose equations are valid, and in which the Ricci tensor vanishes, determines the metric functions q and Q in accordance with the Sachs equations in the form (1.1). Such an initial value function must remain a representation of the same space-time manifold after any of the equivalence transformations corresponding to the infinitesimal generators (3.10) and it must remain related to the transformed metric functions by equations of the form (1.1). Some particular functions $\Psi_0$ may be invariant in form under the action on the Sachs equations of a particular sub-group of equivalence transformations, in which case this sub-group will constitute a symmetry group and the system of equations will admit a corresponding Lie symmetry. As demonstrated in section 2 and section 4, the function (2.23) is the only function that meets these invariance criteria. (The infinitesimal generator of the corresponding Lie symmetry is given by (2.26).) All forms of this function, corresponding to different values of the functions of integration, are transformed by the equivalence transformations into forms of the same general function, and therefore remain consistent with the Sachs equations. The combination of equivalence transformations corresponding to the generator (2.26) leaves both the form of the equations and the form $\overline{\Psi}_0$, of the function that appears in the transformed system of equations, unaltered.[1] A space-time corresponding to this function will therefore have special properties in that its system of Sachs equations possesses the associated non-trivial Lie symmetry.

---

[1] Note that we are considering only transformations of $\Psi_0$ within the context of the Sachs equations, in which all the variables are transformed simultaneously. Considered in isolation, the function is not invariant under the group of symmetry transformations that are applicable to the system of equations as a whole.

The geometric interpretation of the Lie symmetry (2.26) is that for any displacement of the origin along the coordinate curve $\{u, \theta, \phi = \text{const}\}$, both the Sachs equations and the function $\Psi_0$, defined at a point on the same coordinate curve, are form-invariant provided the transformation of r corresponding to a particular displacement (a transformation $\bar{r} = r - a$ where a is some constant) is coupled with a specific dilation (corresponding to a transformation $\bar{r} = a\,r$) and/or a specific rotation of the direction $\frac{\partial}{\partial \theta}$ (which entails the inversion $\bar{r} = \frac{r}{1 - a\,r}$), where the relative magnitudes of these coupled elements of the real sub-group of the Poincare group (also called the inhomogeneous Lorentz group iSL (2R)) are given by the "constants" $\alpha, \beta$ and $\gamma$ (more generally, these are functions of $\theta$ defined on a given null hypersurface).

This is consistent with the fact that a general curved axisymmetric space-time, in which the Ricci tensor vanishes and which is also symmetric with respect to the reflection $d\phi \to -d\phi$, is completely specified within a coordinate patch once the direction of the time vector and the dilation of r have been specified at every point along the null geodesics that constitute the coordinate curves. This follows from the fact that in an orthonormal tetrad, constructed at any point along the geodesic, only the direction $\frac{\partial}{\partial \theta}$ and the direction of the time vector can vary under the action of the null rotations that constitute the gauge freedom of the Newman-Penrose equations: the directions $\frac{\partial}{\partial r}$ and $\frac{\partial}{\partial \phi}$ are fixed in an invariant manner by the tangent to the geodesic $\{u, \theta, \phi = \text{const}\}$ and the Killing vector respectively. Once the direction of the time vector has been fixed, after a null rotation, the direction $\frac{\partial}{\partial \theta}$ follows from the fact that it has to be orthogonal to the directions of the other three tetrad vectors.

It remains an open question whether there are any other functions that have the property that they remain a representation of the same space-time and remain consistent with the Sachs equations in the form (1.1) under the action of the group of equivalence transformations corresponding to the infinitesimal generators (3.10). If not, then the initial value function for all space-times with the geometric symmetries under discussion, and in which the Ricci tensor vanishes, must take the form (1.2). The space-times corresponding to this form of $\Psi_0$, which are special in the sense noted above, presumably also have special physical significance. As remarked in the introduction it has been shown [4] that (1.2) reproduces a number of the features that would be expected, in the appropriate limits, of the vacuum portion of a space-time representing a pair of non-rotating masses moving towards a head-on collision. .

**Appendix:   Derivation of the coordinate form of the Sachs equations**

**Co-ordinate system**
We follow Bondi [15] in introducing a co-ordinate system based on a set of null cones generated from the time-like curve followed by a point O that remains on the axis of symmetry. In the words of Bondi, van der Burg and Metzner [16]: "From the axial symmetry the azimuth angle $\phi$ is readily defined in an invariant manner. Suppose now we put a source of light at a point O on the axis of symmetry and surround it by a small sphere on which we can produce the azimuth co-ordinate $\phi$ together with a co-latitude $\theta$ and a time co-ordinate u. We then define the u, $\theta$, $\phi$ co-ordinates of an arbitrary event E to be the u, θ, φ co-ordinates of the event at which the light ray OE intersects the small sphere. In other words, along an outward radial light ray the three co-ordinates u, θ, φ are constant." An affine parameter r, defined up to a linear transformation, can be associated with the null geodesics. We take the affine parameter as the second co-ordinate and we assign numerical references 1,2,3,4 to u, r, θ and φ in that order.

We now follow Newman and Unti [17] in constructing a null tetrad within the co-ordinate neighbourhood of the origin. For the NP equations to be valid, this co-ordinate neighbourhood has to be a spinor space. This means that it must be possible to construct a null tetrad with a unique orientation at each point. The domain of validity U of the analysis therefore is a simply connected neighbourhood that extends outwards from the origin and terminates if the null geodesics, that constitute the co-ordinate curves, intersect one another.

On each null hypersurface, u = const, a vector $\ell_\alpha = u_{,\alpha}$ can be chosen orthogonal to the hypersurface. Taking u as the co-ordinate $x^1$, it follows that

$$\ell_\alpha = \delta^1_\alpha$$

where δ is the Kronecker delta.

Since the hypersurfaces are null, the vectors $\ell_\alpha$ will also be tangent to the null geodesics lying within the surfaces. The tangents to the geodesics are given by:

$$\ell^\mu = \frac{dx^\mu}{dr} = g^{\mu\nu}\ell_\nu = g^{\mu 1} = \delta^\mu_2$$

It follows that $g^{21} = g^{12} = 1$ while all the remaining elements of the first column and of the first row of the metric tensor $g^{\mu\nu}$ are zero. Similarly,

$$\ell_\mu = g_{\mu\nu}\ell^\nu = g_{\mu 2} = \delta^1_\mu$$

Hence $g_{12} = g_{21} = 1$ while all the other elements of the second column and the second row of the metric tensor $g_{\mu\nu}$ are zero.

From the reflection symmetry under the transformation $d\phi \to -d\phi$ it follows that all the off-diagonal elements of the metric tensor $g_{\mu\nu}$ vanish in the last row and the last column.

It follows from this analysis that the metric tensor has to take the form:

$$g_{\mu\nu} = \begin{pmatrix} g_{11} & 1 & g_{13} & 0 \\ 1 & 0 & 0 & 0 \\ g_{13} & 0 & g_{33} & 0 \\ 0 & 0 & 0 & g_{44} \end{pmatrix} \qquad g^{\mu\nu} = \begin{pmatrix} 0 & 1 & 0 & 0 \\ 1 & g^{22} & g^{23} & 0 \\ 0 & g^{23} & g^{33} & 0 \\ 0 & 0 & 0 & g^{44} \end{pmatrix}$$

**Tetrad and Metric Tensor**

A tetrad of four vectors that are normal to one another is constructed at each point of U, using the tangent vectors $\ell^\mu$ as reference. A second null vector $n^\mu$ is chosen so that $\ell^\mu n_\mu = 1$. The tetrad is then completed with a pair of complex null vectors, $m^\mu$ and $\overline{m}^\mu$, that are normal to each other and to the pair of real null vectors. These complex null vectors are defined from an orthonormal pair of real unit space-like vectors, $a^\mu$ and $b^\mu$ such that

$$m^\mu = \tfrac{1}{\sqrt{2}}(a^\mu + i b^\mu) \quad \text{and} \quad \overline{m}^\mu = \tfrac{1}{\sqrt{2}}(a^\mu - i b^\mu)$$

The pair-wise inner products of the tetrad vectors are determined by their normality conditions:

$$\ell_\mu \ell^\mu = m_\mu m^\mu = \overline{m}_\mu \overline{m}^\mu = n_\mu n^\mu = 0$$

$$\ell_\mu n^\mu = -m_\mu \overline{m}^\mu = 1$$

$$\ell_\mu m^\mu = \ell_\mu \overline{m}^\mu = n_\mu m^\mu = n_\mu \overline{m}^\mu = 0$$

The tetrad vectors are related to the elements of the metric tensor by the relation

$$g^{\mu\nu} = \ell^\mu n^\nu + n^\mu \ell^\nu - m^\mu \overline{m}^\nu - \overline{m}^\mu m^\nu$$

A similar relation holds for the covariant form of the metric tensor.

To satisfy the normality conditions the vectors $n^\mu$ and $m^\mu$ must take the form

$$n^\mu = \delta_1^\mu + U \delta_2^\mu + X^i \delta_i^\mu \qquad (i = 3, 4)$$

$$m^\mu = \omega \delta_2^\mu + \xi^i \delta_i^\mu$$

where, using the notation of Newman, Penrose, and their co-workers, U, $X^i$, $\omega$ and $\xi^i$ are unknown, complex-valued functions of u, r, and θ. Using these relations gives the elements of the metric tensor $g^{\mu\nu}$ in terms of the functions that appear in the tetrad vectors.

It is possible to choose the vector $n^\mu$ such that its spatial projection lies in the r–θ plane at the origin. This implies $X^4 = 0$. As a result of the axial symmetry, the spatial projection of $n^\mu$ must then remain in the r–θ plane as the tetrad propagates outwards, if it is parallel propagated, so that $X^4 = 0$ for all r. The index is then redundant and it is possible to put $X^3 \equiv X$.

If the imaginary axis is chosen to lie along the co-ordinate direction $x^4$ and one puts $\xi^3 = \dfrac{1}{q}$ and $\xi^4 = \dfrac{i}{Q}$ where both q and Q are real-valued functions of the co-ordinates the result is that all the metric functions become real-valued as do all the Newman-Penrose equations and all the functions that appear in them. The metric tensor resulting from this analysis is:

$$ds^2 = (-2U - \tfrac{1}{2}X^2q^2 + 2X\omega q)du^2 + 2dudr - (2\omega q - q^2 X)dud\theta - \tfrac{1}{2}q^2 d\theta^2 - \tfrac{1}{2}Q^2 d\phi^2$$

$$g^{\mu\nu} = \begin{pmatrix} 0 & 1 & 0 & 0 \\ 1 & 2(U-\omega^2) & X - \dfrac{2\omega}{q} & 0 \\ 0 & X - \dfrac{2\omega}{q} & \dfrac{-2}{q^2} & 0 \\ 0 & 0 & 0 & \dfrac{-2}{Q^2} \end{pmatrix}$$

There is an apparent redundancy since five functions are used to specify four elements of the metric tensor. However, when parallel propagation is imposed as a condition, a relation between ω and X results, which removes the redundancy (if the tetrad is parallel propagated then the spin coefficient π must vanish).

**Transformations of the co-ordinates**
Two sets of transformations can be applied to the tetrad without changing the metric tensor, violating the normality conditions, or changing the direction of the tangent vector $\ell^\mu$:

- Rotation of the pair of space-like vectors in their plane without changing their orientation relative to each other, i.e.

$$\ell^{\mu'} = \ell^\mu, \qquad n^{\mu'} = n^\mu, \qquad m^{\mu'} = m^\mu e^{iC} \quad \text{(C real)}$$

Null rotations that leave $\ell^\mu$ fixed, i.e.

$$\ell^{\mu'} = \ell^\mu, \qquad n^{\mu'} = n^\mu + \overline{B} m^\mu + B \overline{m}^\mu + B\overline{B}\ell^\mu, \qquad m^{\mu'} = m^\mu + B\ell^\mu$$

In general the parameter B is complex. However, if the condition is imposed that the imaginary space-like direction must remain aligned with the Killing field $\dfrac{\partial}{\partial \phi}$, then B becomes real. Under this condition the parameter C must also vanish, leaving only the real null rotations as a gauge freedom.

**The Sachs equations**
In the notation of Newman and Penrose, the two Sachs equations take the following form in the current case where the Ricci tensor vanishes and all the variables are real-valued:

$$\frac{\partial \rho}{\partial r} = \rho^2 + \sigma^2 \quad \text{and} \quad \frac{\partial \sigma}{\partial r} = 2\rho\sigma + \Psi_0$$

where the function $\Psi_0$ is the first term of the Weyl spinor and ρ and σ, the so-called optical scalars, are the expansion and shear, respectively, of the null geodesic congruence. They are defined by:

$$\rho = \ell_{\mu;\nu} m^\mu \overline{m}^\nu \quad \text{and} \quad \sigma = \ell_{\mu;\nu} m^\mu m^\nu$$

Expanding the expressions for ρ and σ in terms of the above metric tensor yields the equations in the form (1.1).